\documentclass[aps,prb,twocolumn]{revtex4}
\usepackage{graphicx}
\usepackage{amsmath}
\usepackage{graphics}
\usepackage{epsfig}
\usepackage{yfonts}
\usepackage{amsfonts}
\usepackage{amssymb}
\usepackage{xcolor,cancel}
\usepackage{bm}
\usepackage{float}
\usepackage{caption}
\usepackage{subcaption}
\usepackage{tikz}
\usepackage{tikz-feynman}
\usepackage{feynmp-auto}
\usepackage{pgfplots}
\usepackage{adjustbox}

\usetikzlibrary{decorations.pathmorphing}

\newcommand{\overbar}[1]{\mkern 1.5mu\overline{\mkern-1.5mu#1\mkern-1.5mu}\mkern 1.5mu}

\begin{document}

\title{Floquet Nonequilibrium Green's functions with Fluctuation-Exchange Approximation: Application to Periodically Driven Capacitively Coupled Quantum Dots}

\author{Thomas D. Honeychurch} 
\author{Daniel S. Kosov}
\affiliation{College of Science and Engineering, James Cook University, Townsville, QLD, 4811, Australia}

\begin{abstract}
We study the dynamics of two capacitively coupled quantum dots, each coupled to a lead. A Floquet Green's function approach described the system's dynamics, with the electron-electron interactions handled with the fluctuation-exchange approximation. While electrons cannot move between the separate sections of the device, energy transfer occurs with the periodic driving of one of the leads. This process was found to be explained with four stages. The energy transfer was also found to be sensitive to the driving frequency of the leads, with an optimal frequency corresponding to the optimal completion of the four stages of the identified process.
\end{abstract}

\maketitle

\section{Introduction}

Capacitive coupling offers a unique tool for investigating and designing open quantum systems. Of particular interest are systems where energy transport occurs between regions without the addition of charge transport. Interesting phenomena that utilize capacitive coupling include coulomb drag\cite{Keller2016,Sierra2019} and heat rectification\cite{Tesser2022}. Capacitively coupled quantum dots offer a simple testbed for such phenomena, with heat current across capacitively coupled quantum dots\cite{Arrachea2020,Harbola2019} and their use in energy harvesting devices \cite{Dare2019,Sanchez2011,Thierschmann2015,Sothmann2015,Kaasbjerg2017} having been investigated.

The addition of time-dependent drivings of lead and gate voltages offers a further avenue for exploring particle and energy transport within quantum devices, most usually in the case of periodic drivings\cite{Ludovico2016_review}: energy transport and entropy production of a noninteracting single electronic level with a periodically modulated gate voltage has been discussed \cite{Ludovico2016,Ludovico2014}; the periodic modulation of parameters has also been utilized to investigate nanoscale thermal machines\cite{Dare2016,Ludovico2018,Juergens2013}; and the AC linear response of both particle and heat current has been investigated for a mesoscopic capacitor\cite{Sanchez2013}. In the context of capacitively coupled devices, the electrothermal admittance has been calculated for a nanoscale parallel plate capacitor in the linear response regime\cite{Chen2015}, and, most recently, the energy transfer in a system of capacitively coupled dots was investigated when the gate voltage of one dot is modulated periodically\cite{Ludovico2022}. 

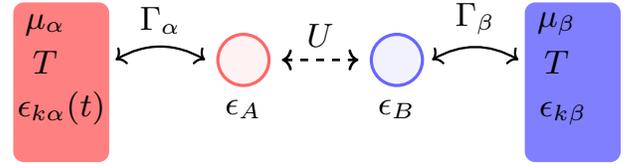
\begin{figure}[h!]
			\hspace*{0cm}
	\begin{tikzpicture}[scale=0.85, every node/.style={scale=1.4}]

		\fill[color=red!60, fill=red!50,rounded corners] (0,0) rectangle (1.5,2.5);
		
		\node[above,color=white,text=black] at (0.5,1.8) {$\mu_\alpha$};
		\node[above,color=white,text=black] at (0.5,1.2) {$T$};
		\node[above,color=white,text=black] at (0.75,0.4) {$\epsilon_{k\alpha} (t)$};

		\fill[color=blue!60, fill=blue!50,rounded corners] (8,0) rectangle (9.5,2.5);
		
		\node[above,color=white,text=black] at (8.5,1.8) {$\mu_\beta$};
		\node[above,color=white,text=black] at (8.5,1.2) {$T$};
		
		\draw[<->,line width=0.3mm] (1.6,1.6) to [bend left] node[midway,above] {$\Gamma_\alpha$}  (3,1.6);
		\draw[<->,line width=0.3mm] (6.55,1.6) to [bend left] node[midway,above] {$\Gamma_\beta$}  (7.9,1.6);
		\node[above,color=white,text=black] at (8.6,0.4) {$\epsilon_{k\beta}$};
		
		\filldraw[color=red!60, fill=red!5, very thick] (3.6,1.6) circle (0.4); 
		\node[above,color=white,text=black] at (3.6,0.5) {$\epsilon_A$};
		
		\filldraw[color=blue!60, fill=blue!5, very thick] (6,1.6) circle (0.4); 
		\node[above,color=white,text=black] at (6,0.5) {$\epsilon_B$};
		
		\draw [<->,line width=0.35mm,dashed] (4.2,1.6) -- (5.4,1.6) node[midway,above] {$U$};
	
	\end{tikzpicture}
		\caption{Schematic representation of the model investigated. The two quantum dots are coupled to noninteracting electron reservoirs and coupled to each other by Coulomb interaction. Within the investigation, energies of reservoir $A$ are driven harmonically, resulting in a nonzero current between the dots and reservoirs and energy transfer between the reservoirs.\label{fig:schematic}}
\end{figure}

This paper investigates the energy transfer between two capacitively coupled quantum dots, each connected to a respective lead [see Fig. \ref{fig:schematic}]. We study the energy and particle transport within the system due to the periodic driving of one lead's energies. While particles cannot move between systems, the capacitive coupling between the dots allows energy transfer through the system. We make use of a Floquet nonequilibrium Green's functions approach\cite{Brandes1997,Honeychurch2020,Honeychurch2023,Haughian2017,Aoki2014}, allowing for the exploration of nonadiabatic drivings. The Coulomb interaction is handled with self-consistent perturbation theory, using the fluctuation-exchange (FLEX) approximation\cite{Schlunzen2020}. A self-consistent approximation, FLEX includes both particle-particle and particle-hole T-matrix and GW terms [see Fig. \ref{fig:diagrams}]. FLEX subsumes the advantages of its constituent terms, making it applicable to a wide variety of interaction strengths and occupations\cite{Schlunzen2020}.

It was found that the average energy current through the system is sensitive to the driving frequency, with a frequency corresponding to the maximum energy transference observed. This energy transfer was found to be described by a four-stage process. The effects of the other parameters were also investigated.

The paper is organized as follows: Section \ref{sec:theory} lays out theory and implementation; Section \ref{sec:discussion} investigates the energy transfer while one lead is driven periodically; and within section \ref{ref:conclusion} the paper's results are summarized and extensions suggested. Natural units for quantum transport are used throughout the paper, with $\hbar$, $e$, and $k_B$ set to unity. 

\section{Theory} \label{sec:theory}

\subsection{Hamiltonian and NEGF}

For simplicity, we focus on two spinless dots, $A$ and $B$, coupled to an associated lead, labeled $\alpha$ and $\beta$, and coupled capacitively:
\begin{equation}
\begin{split}
	H (t) = H_A + H_B + H_{int} + H_{A\alpha} + H_{B\beta} \\ + H_\alpha (t) + H_\beta,
\end{split}
\end{equation}
\begin{equation}
	H_S  = \epsilon_S \hat{d}_S^\dagger \hat{d}_S, \;\;\;\;\;\;\;\;\; H_{int} = U \hat{d}_A^\dagger \hat{d}_A \hat{d}_B^\dagger \hat{d}_B,
\end{equation}
\begin{equation}
	H_{S\sigma} = \sum_{k} t_{k\sigma S} \; \hat{c}^{\dagger}_{k \sigma} \hat{d}_S + t^*_{k\sigma S} \; \hat{d}^\dagger_S \hat{c}_{k \sigma},
\end{equation}
and
\begin{equation}
	H_{\sigma} = \sum_{k}  \left(\epsilon_{k\sigma} + \psi_\sigma \left(t\right)\right) \hat{c}^{\dagger}_{k\sigma} \hat{c}_{k\sigma}.
\end{equation}
Here, $S$ refers to a dot, $\sigma$ refers to its corresponding lead, and $\bar S$ and $\bar \sigma$ refer to the opposing dot and lead, respectively. The two interacting dots' energies are given by $\epsilon_S$, and the electron-electron repulsion between the sites, given by $H_{int}$, has a strength $U$. The coupling of the quantum dots to their respective leads is governed by $H_{k\sigma S}$, with $t_{k\sigma S}$ denoting a hopping between the lead site $k\sigma$ and the dot $S$. The leads are taken as noninteracting, with the explicit time-dependence entering the Hamiltonian via $\psi_\sigma (t)$, which varies the energies $\epsilon_{k\sigma}$.

\begin{figure}[ht!]
	\raggedright
	\begin{subfigure}[c]{1in}
		\begin{align*}
			\Sigma^{SB}_S \; = \quad
			\begin{tikzpicture}[scale=1, every node/.style={scale=1},baseline=0.5cm]
				\begin{feynman}[]
					\vertex (a);
					\vertex [right = 0.8cm of a] (b);
					\vertex [above = 0.8cm of a] (t1);
					\vertex [above = 0.8cm of b] (t2);
					\diagram* {
						(b) -- [red,thick,fermion] (a),
						(a) -- [dashed] (t1),
						(b) -- [dashed] (t2),
						(t1) -- [blue,thick,fermion,quarter left] (t2) -- [blue,thick,fermion,quarter left] (t1),
					};
				\end{feynman}
			\end{tikzpicture} 
		\end{align*}
	\end{subfigure}
	\hfill
	\begin{subfigure}[c]{1in}
		\begin{align*}
			\Sigma^{GW}_S \; = \quad
			\begin{tikzpicture}[scale=1, every node/.style={scale=1},baseline=0.5cm]
				\begin{feynman}
					\vertex (a);
					\vertex [right = 0.8cm of a] (b);
					\vertex [above = 0.8cm of a] (t1);
					\vertex [above = 0.8cm of b] (t2);
					\diagram* {
						(b) -- [red,thick,fermion] (a),
						(a) -- [dashed] (t1),
						(b) -- [dashed] (t2),
						(t1) -- [blue,thick,fermion,quarter left] (t2) -- [blue,thick,fermion,quarter left] (t1),
					};
				\end{feynman}
			\end{tikzpicture} 
			\quad + \quad
			\begin{tikzpicture}[scale=1, every node/.style={scale=1},baseline=0.5cm]
				\begin{feynman}
					\vertex (a);
					\vertex [right = 3.6cm of a] (b);
					\vertex [above = 0.8cm of a] (t1);
					\vertex [above right = 0.8*1.41421cm of a] (t2);
					\vertex [right = 0.6*1cm of t2] (t3);
					\vertex [right = 0.8*1cm of t3] (t4);
					\vertex [right = 0.6*1cm of t4] (t5);
					\vertex [right = 0.8*1cm of t5] (t6);
					\diagram* {
						(b) -- [red,thick,fermion] (a),
						(a) -- [dashed] (t1),
						(b) -- [dashed] (t6),
						(t2) -- [dashed] (t3),
						(t4) -- [dashed] (t5),
						(t1) -- [blue,thick,fermion,quarter left] (t2) -- [blue,thick,fermion,quarter left] (t1),
						(t5) -- [blue,thick,fermion,quarter left] (t6) -- [blue,thick,fermion,quarter left] (t5),
						(t3) -- [red,thick,fermion,quarter left] (t4) -- [red,thick,fermion,quarter left] (t3),
					};
				\end{feynman}
			\end{tikzpicture} 
			\quad + \; \cdots \quad
		\end{align*}
	\end{subfigure}
	
	\begin{subfigure}[c]{1in}
		\begin{align*}
			\Sigma^{TPP}_S \; = \quad
			\begin{tikzpicture}[scale=1, every node/.style={scale=1},baseline=0.5cm]
				\begin{feynman}
					\vertex (a);
					\vertex [right = 0.8cm of a] (b);
					\vertex [above = 0.8cm of a] (t1);
					\vertex [above = 0.8cm of b] (t2);
					\diagram* {
						(b) -- [red,thick,fermion] (a),
						(a) -- [dashed] (t1),
						(b) -- [dashed] (t2),
						(t1) -- [blue,thick,fermion,quarter left] (t2) -- [blue,thick,fermion,quarter left] (t1),
					};
				\end{feynman}
			\end{tikzpicture} 
			\quad + \quad
			\begin{tikzpicture}[scale=1, every node/.style={scale=1},baseline=0.5cm]
				\begin{feynman}
					\vertex (a);
					\vertex [right = 0.8cm of a] (b1);
					\vertex [right = 1.6cm of a] (b);
					\vertex [above = 0.8cm of a] (t1);
					\vertex [right = 0.8cm of t1] (t2);
					\vertex [right = 0.8cm of t2] (t3);
					\diagram* {
						(b) -- [red,thick,fermion] (b1),
						(b1) -- [red,thick,fermion] (a),
						(b1) -- [dashed] (t2),
						(a) -- [dashed] (t1),
						(b) -- [dashed] (t3),
						(t1) -- [blue,thick,fermion,quarter left] (t3),
						(t2) -- [blue,thick,fermion,quarter left] (t1),
						(t3) -- [blue,thick,fermion,quarter left] (t2),
					};
				\end{feynman}
			\end{tikzpicture} 
			\quad + \; \cdots \quad
		\end{align*}
	\end{subfigure}
	
	\begin{subfigure}[c]{1in}
		\begin{align*}
			\Sigma^{TPH}_S \; = \quad
			\begin{tikzpicture}[scale=1, every node/.style={scale=1},baseline=0.5cm]
				\begin{feynman}
					\vertex (a);
					\vertex [right = 0.8cm of a] (b);
					\vertex [above = 0.8cm of a] (t1);
					\vertex [above = 0.8cm of b] (t2);
					\diagram* {
						(b) -- [red,thick,fermion] (a),
						(a) -- [dashed] (t1),
						(b) -- [dashed] (t2),
						(t1) -- [blue,thick,fermion,quarter left] (t2) -- [blue,thick,fermion,quarter left] (t1),
					};
				\end{feynman}
			\end{tikzpicture} 
			\quad + \quad
			\begin{tikzpicture}[scale=1, every node/.style={scale=1},baseline=0.5cm]
				\begin{feynman}
					\vertex (a);
					\vertex [right = 0.8cm of a] (b1);
					\vertex [right = 1.6cm of a] (b);
					\vertex [above = 0.8cm of a] (t1);
					\vertex [right = 0.8cm of t1] (t2);
					\vertex [right = 0.8cm of t2] (t3);
					\diagram* {
						(b) -- [red,thick,fermion] (b1),
						(b1) -- [red,thick,fermion] (a),
						(b1) -- [dashed] (t2),
						(a) -- [dashed] (t1),
						(b) -- [dashed] (t3),
						(t3) -- [blue,thick,fermion,quarter right] (t1),
						(t1) -- [blue,thick,fermion,quarter right] (t2),
						(t2) -- [blue,thick,fermion,quarter right] (t3),
					};
				\end{feynman}
			\end{tikzpicture} 
			\quad + \; \cdots \quad
		\end{align*}
	\end{subfigure}
	
	\caption{The Feynman diagrams considered within the investigation. Here, $S$ refers to the red fermionic line and corresponds to $G_S(\tau,\tau')$. The blue fermionic line corresponds to the opposing dot's Green's function, $G_{\bar S}(\tau,\tau')$.\label{fig:diagrams}}
\end{figure}
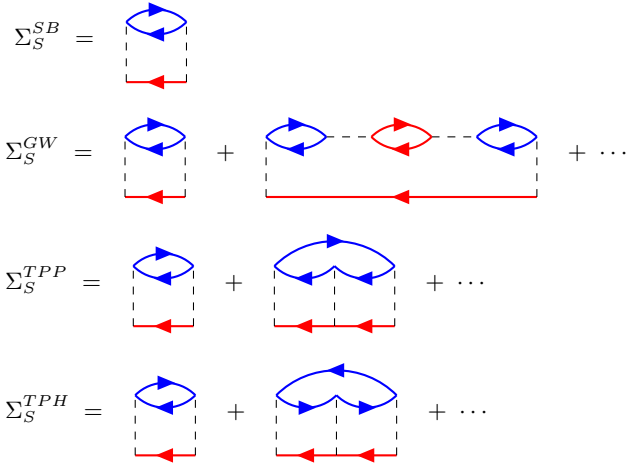

To model the system out of equilibrium, we make use of nonequilibrium Green's functions:
\begin{equation}
	G_{S} (\tau,\tau') = -i\left\langle T_c\left(d_S \left(\tau\right)d^\dagger_S \left(\tau'\right)\right) \right\rangle,
\end{equation}
with the equation of motion,
\begin{equation}
	\begin{split}
	 \left(i\frac{\partial}{\partial \tau} -\epsilon_S \right)G_{S} \left(\tau,\tau'\right) \\- \int_c d\tau_1 \; \Sigma_{S} \left(\tau,\tau_1\right) G_{S} \left(\tau_1,\tau'\right) =  \delta_c \left(\tau-\tau'\right),
	\end{split}
\end{equation}
where the self-energy term consists of contributions from the associated lead and the interaction between the quantum dots:
\begin{equation}
	\Sigma_{S} \left(\tau,\tau'\right) = \Sigma_{\sigma} \left(\tau,\tau'\right) + \Sigma^{int}_{S} \left(\tau,\tau'\right).
\end{equation}

To account for the capacitive coupling between the dots, we make use of self-consistent perturbation theory:
\begin{equation}
	\Sigma^{int}_S \left(\tau,\tau'\right) = -i U G_{\bar{S}} \left(\tau,\tau^+\right) \delta(\tau,\tau') + \Sigma^{corr}_S (\tau,\tau'),
\end{equation}
where the correlations were investigated with the FLEX approximation\cite{Schlunzen2020}:
\begin{equation}
	\begin{split}
		\Sigma^{FLEX}_{S} (\tau,\tau') = \Sigma^{TPP}_{S} (\tau,\tau') + \Sigma^{TPH}_{S} (\tau,\tau') \\  + \Sigma^{GW}_{S} (\tau,\tau')  - 2 \Sigma^{SB}_{S} (\tau,\tau').
	\end{split}
\end{equation}
The single-bubble, GW, particle-particle, and particle-hole T-matrix approximations follow standard definitions here [see Fig. \ref{fig:diagrams}]. The single-bubble approximation is given by
\begin{equation}
	\Sigma^{SB}_S \left(\tau,\tau'\right) = U G_{\bar{S}} \left(\tau,\tau'\right) G_{\bar{S}} \left(\tau',\tau\right) U G_S \left(\tau,\tau'\right).
\end{equation}
The $GW$ self-energy is given by 
\begin{equation}
	\begin{split}
		\Sigma^{GW}_{S} (\tau,\tau') = 
		i W_S^{ns}\left(\tau,\tau'\right) G_S \left(\tau,\tau'\right),
	\end{split}
\end{equation}
\begin{equation}
	\begin{split}
		W^{ns}_S \left(\tau,\tau'\right) =   \Phi_{S} \left(\tau,\tau'\right) 
		\\ + \int_c \tau_1 \int_c \tau_2 \;   \Phi_{S} \left(\tau,\tau_1\right)  P_S \left(\tau_1,\tau_2\right) W_S^{ns} \left(\tau_2,\tau'\right), 
	\end{split}
\end{equation}
\begin{equation}
	\begin{split}
		\Phi_S \left(\tau,\tau'\right) = U P_{\overbar S} \left(\tau,\tau'\right) U
	\end{split}
\end{equation}
and
\begin{equation}
	\begin{split}
		P_S \left(\tau,\tau'\right) = -i G_S \left(\tau,\tau'\right) G_S \left(\tau',\tau\right).
	\end{split}
\end{equation}
The particle-particle T-matrix self-energy is given by
\begin{equation}
	\Sigma^{PP}_S \left(\tau,\tau'\right) = i \; T^{PP} (\tau,\tau') G_{\bar S} \left(\tau',\tau\right),
\end{equation}
\begin{equation}
	\begin{split}
		T^{PP} \left(\tau,\tau'\right) = - U G^H \left(\tau,\tau'\right) U
		\\
		+ \int d\tau_1 U G^H \left(\tau,\tau_1\right) T^{PP} \left(\tau_1,\tau'\right)
	\end{split}
\end{equation}
and
\begin{equation}
	\begin{split}
		G^H (\tau,\tau') = iG_A (\tau,\tau') G_B (\tau,\tau').
	\end{split}
\end{equation}
The particle-hole T-matrix self-energy is given by 
\begin{equation}
	\Sigma^{PH}_S \left(\tau,\tau'\right) = i \; T_{S}^{PH} (\tau,\tau') G_{\overbar S} \left(\tau,\tau'\right),
\end{equation}
\begin{equation}
	\begin{split}
		T_{S}^{PH} \left(\tau,\tau'\right) = U G_{S\overbar S}^F \left(\tau,\tau'\right) U
		\\
		- \int d\tau_1 U G_{S\overbar S}^F \left(\tau,\tau_1\right) T_{S}^{PH} \left(\tau_1,\tau'\right)
	\end{split}
\end{equation}
and
\begin{equation}
	\begin{split}
		G^F_{S\bar S} (\tau,\tau') = -iG_S (\tau,\tau') G_{\bar S} (\tau',\tau).
	\end{split}
\end{equation}

For the leads, time dependence within the energies, $\epsilon_{k\sigma} + \psi_\sigma(t)$, results in an additional phase to the otherwise static lead self-energies:
\begin{equation} \label{eq:lead_self_energies}
	\begin{split}
		\Sigma_{\sigma} (t,t') 
		= \Sigma'_{\sigma} (t - t') e^{-i \int_{t'}^{t} dt_1 \psi_{\sigma}(t_1)}
		\\=  e^{-i\Psi_{\sigma}(t)} \Sigma'_{\sigma} (t - t') e^{i\Psi_{\sigma}(t')},
	\end{split}
\end{equation}
where $\Psi_{\sigma}(t)$ is the anti-derivative of $\psi_\sigma(t)$ and $\Sigma'_{\sigma} (t - t')$ is the self-energies of the leads without driving, given in Fourier space with the wide-band approximation as 
\begin{equation}
	\begin{split}
		\Sigma_{\sigma}^{R/A} \left(\omega\right) = \mp \frac{i}{2} \Gamma_{\sigma},
	\end{split}
\end{equation}
\begin{equation}
	\begin{split}
		\Sigma^{<}_{\sigma} \left(\omega\right) = i f_\sigma \left(\omega\right) \Gamma_{\sigma},
	\end{split}
\end{equation}
\begin{equation}
	\begin{split}
		\Sigma^{>}_{\sigma} \left(\omega\right) = -i\left(1 - f_\sigma \left(\omega\right)\right) \Gamma_{\sigma},
	\end{split}
\end{equation}
where the Fermi distribution follows the standard definition of $f_\sigma (\omega) = 1 / \left[\exp\left(\left(\omega - \mu_\sigma\right)/T\right) + 1\right]$. 

The particle current is given by,
\begin{equation}
	\begin{split}\label{eq:current}
		I^P_{\sigma} (t) =  2 Re \left\{\int^{\infty}_{-\infty} dt_1 Tr\left[G_S^<(t,t_1) \Sigma_{\sigma}^A (t_1,t) \right.\right.
		\\
		\left.\left. + G_S^R (t,t_1) \Sigma_{\sigma}^< (t_1,t)\right]\right\},
	\end{split}
\end{equation}
and the occupation of the dots is given by 
\begin{equation}\label{eq:electronic_occupation}
	n_{S} (t) = -i G_{S}^<\left(t,t\right),
\end{equation}
where continuity dictates that $I^P_{\sigma} (t) = - \frac{d n_S (t)}{dt}$. To calculate the energy that passes from the leads into the system, we use
\begin{equation}
	\begin{split}
		I^E_\sigma (t) = -i \langle \left[H(t),H_{\sigma} (t)\right]_- \rangle 
		\\
		= -2 \operatorname{Re} \left\{\int dt_1 \left[ i\frac{d}{dt} \Sigma^<_{\sigma} (t,t_1) \right] G^A_{S} (t_1,t) \right. \\ \left. + \left[ i\frac{d}{dt}\Sigma^R_{\sigma} (t,t_1) \right] G^<_{S} (t_1,t)  \right\}.
	\end{split}
\end{equation}
Here, energy moves from the leads into the system and system-lead coupling, resulting in the continuity equation
\begin{equation}\label{eq:time_resolved_energy_continuity}
	\begin{split}
		I^E_A (t) + I^E_B (t) = \frac{d}{dt} \left( \langle H_{A\alpha} \rangle +  \langle H_{B\beta} \rangle \right. \\  \left. + \langle H_{A} \rangle + \langle H_{B} \rangle +  E_{int} (t) \right)
	\end{split}
\end{equation}
where
\begin{equation}
	\begin{split}
		\langle H_{S\sigma} \rangle = 2 \operatorname{Im} \left\{\int dt_1 \Sigma^<_{\sigma} (t,t_1) G^A_{S} (t_1,t) \right. \\ + \left. \Sigma^R_{\sigma} (t,t_1) G^<_{S} (t_1,t) \right\}
	\end{split}
\end{equation}
and
\begin{equation}
	\begin{split}
		E_{int} (t) = \sum_{S = A,B} -\frac{i}{2} \left[ \int dt_1 \; \Sigma^{int,R}_{S} \left(t,t_1\right) G^<_{S} \left(t_1,t\right) \right. \\ \left. +  \Sigma^{int,<}_{S} \left(t,t_1\right) G^A_{S} \left(t_1,t\right) \right].
	\end{split}
\end{equation}
Taking the time-average of equation (\ref{eq:time_resolved_energy_continuity}), gives 
\begin{equation}
	\bar{I}^E_A  = - \bar{I}^E_B,
\end{equation}
where $\bar O = \lim_{\tau \rightarrow \infty} \left(\int^\tau_0  O (t) \;dt\right) / \tau $.

\subsection{Floquet approach}

We use a Floquet nonequilibrium Green's function approach to solve the equations of motion, assuming the periodicity of the system's dynamics\cite{Brandes1997,Honeychurch2020,Honeychurch2023,Haughian2017,Aoki2014}. Within this context, Green's functions are periodic in the central time, $T=\frac{t+t'}{2}$: 
\begin{equation}\label{eq:periodicity_1}
	\begin{split}
		A(t,t') = A\left(T=\frac{t+t'}{2},\tau=t-t'\right) \\ = \sum^{\infty}_{n= -\infty} A(\tau,n) e^{i\Omega nT}
	\end{split}
\end{equation}
and
\begin{equation}\label{eq:periodicity_2}
	A\left(\omega,n\right) = \frac{1}{P} \int^P_{0} dT \; e^{-i\Omega n T} \int^{\infty}_{-\infty} d\tau e^{i\omega \tau} A(T,\tau),
\end{equation} 
which allows us to cast the terms $C_+(t,t') = A(t,t')B(t,t')$ and
$C_{-}(t,t') = A(t,t')B(t',t)$ as
\begin{equation}	
	\begin{split}
		C_{\pm}\left(\omega,n\right)= \\
		\sum^\infty_{m=-\infty} \; \int \frac{d\omega'}{2\pi}	
		A\left(\omega',m\right) B 	\left(\pm\omega\mp\omega',n-m\right)
		\\
		= \left[A \; \square_{\pm} \; B\right]
		\left(\omega,n\right).
	\end{split}
\end{equation}
Making a further transformation of the two-time objects into the Floquet matrix form 
\begin{equation} \label{eqn:floquet_matrix}
	\widetilde{A} \left(\omega,m,n\right) = A \left(\omega + \frac{\Omega}{2} \left(m + n\right), n-m\right)
\end{equation}
allows for convolutions of the type $C(t,t') = \int dt_1 A(t,t_1) B(t_1,t')$ to be recast as matrix equations
\begin{equation} \label{eq:floquet_convolution}
	\begin{split}
		\widetilde{C} \left(\omega,m,n\right)  = \sum_{r=-\infty}^{\infty} \widetilde{A} \left(\omega,m,r\right) \widetilde{B} \left(\omega,r,n\right)
		\\
		= \left[\widetilde{A} \circ \widetilde{B}\right] \left(\omega,m,n\right).
	\end{split}
\end{equation}

\begin{figure*}[pt!]
	\centering
	\hspace*{-4.5cm} 
	\begin{subfigure}[]{1in}
		\centering
		\includegraphics[width=2.5\textwidth]{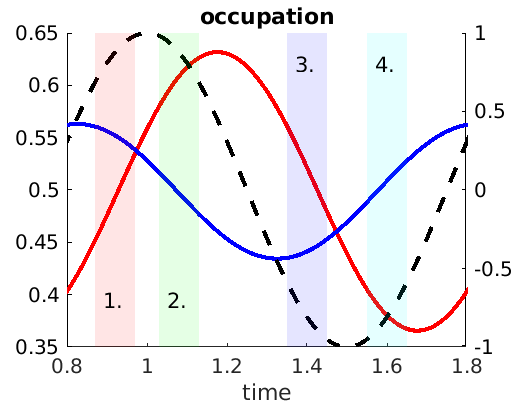} 
		\caption{\label{fig:plotA1}}
	\end{subfigure}
	\qquad\qquad\qquad\qquad\qquad\qquad
	\begin{subfigure}[]{1in}
		\centering
		\includegraphics[width=2.5\textwidth]{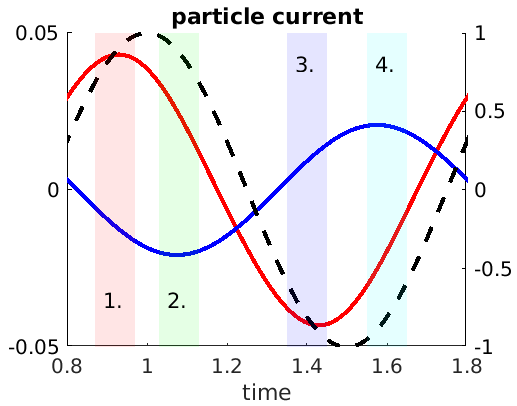} 
		\caption{\label{fig:plotA2}}
	\end{subfigure}
	\qquad\qquad\qquad\qquad\qquad\qquad
	\begin{subfigure}[]{1in}
		\centering
		\includegraphics[width=2.5\textwidth]{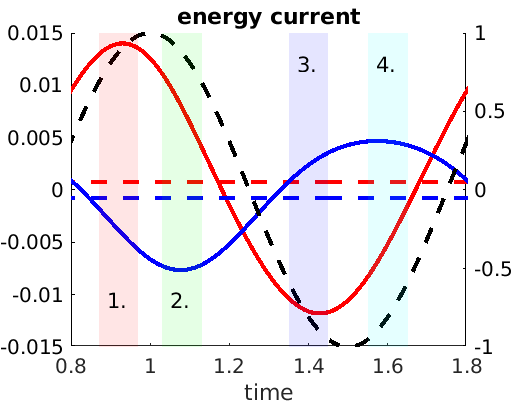} 
		\caption{\label{fig:plotA3}}
	\end{subfigure}
	
	\caption{Observables measured over a period with driving of the left lead. Here, red lines correspond to the observables relating to the driven section, while blue lines refer to the undriven section of the model. Here, the energy current refers to the energy transfer from the lead and coupling region into the central region. The black dashed line is given by $\cos(\Omega t)$, while the colored dashed lines of Fig. \ref{fig:plotA3} correspond to the averages of similarly colored energy currents. The parameters are $\Gamma_\alpha =\Gamma_\beta =0.5$, $U=0.6$, $T=0.001$, $\mu_\alpha =\mu_\beta =0.3$, $\epsilon_A = \epsilon_B =0$, $\Delta_\alpha=0.2$ and $\Omega = 0.32$. The discretization was taken at $0.01$, the bounds of integration between $-40$ and $40$, and 49 Fourier coefficients were used. The convergence for both dots was taken as $10^{-4}$.}
	\label{fig:plotA}
\end{figure*}

Taking real-time projections, we transform the equations of motion into matrix equations with the above transformations:
\begin{equation}\label{eqn:KB-R/A}
	\begin{split}
		\left(\omega + \Omega m - \epsilon_{S} \right)\widetilde{G}_{S}^{R/A} \left(\omega,m,n\right) = 
		\\
		\delta_{mn} + \left[\widetilde{\Sigma}_{S}^{R/A} \circ \widetilde{G}_{S}^{R/A}\right] \left(\omega,m,n\right),
	\end{split}
\end{equation}
\begin{equation}
	\begin{split}
		\widetilde{G}_{S}^{<} (\omega,m,n) =
		\\
		\left[\widetilde{G}_{S}^R \circ \widetilde{\Sigma}^<_{S} \circ \widetilde{G}_{S}^A \right] \left(\omega,m,n\right).
	\end{split}
\end{equation}
The interaction self-energies can be cast in terms of Fourier coefficients: for GW, 
\begin{equation}\label{eq:se_GW_R/A}
	\begin{split}
		\Sigma^{R/A}_{GW,S} \left(\omega,n\right) \\ = i\left[W_S^{ns,<} \;\square_+ \; G^{R/A}_S + W_S^{ns,R/A} \;\square_+ \; G^{>}_S \right]\left(\omega,n\right),
	\end{split}
\end{equation}
\begin{equation}\label{eq:se_GW_</>}
	\Sigma^{</>}_{GW,S} \left(\omega,n\right) = i\left[W_S^{ns,</>} \;\square_+ \; G^{</>}_S \right]\left(\omega,n\right),
\end{equation}
\begin{equation}\label{eq:P_s_R/A}
	\begin{split}
		P_S^{R/A} (\omega,n) \\
		= -i\left[G_S^{<} \;\square_- \; G^{A/R}_S + G_S^{R/A} \;\square_- \; G^{<}_S \right] (\omega,n),
	\end{split}
\end{equation}
\begin{equation}\label{eq:P_s_</>}
	\begin{split}
		P_S^{</>} (\omega,n) \\
		= -i\left[G_S^{</>} \;\square_- \; G^{>/<}_{S} \right] (\omega,n),
	\end{split}
\end{equation}
\begin{equation}\label{eqn:wns_R/A}
	\begin{split}
		\widetilde{W}_S^{ns,R/A} \left(\omega,m,n\right) = \widetilde{\Phi}^{R/A}_S \left(\omega,m,n\right) 
		\\ +  \left[\widetilde{\Phi}^{R/A}_S \circ  \widetilde{P}^{R/A}_S   \circ \widetilde{W}_S^{ns,R/A} \right] \left(\omega,m,n\right)
	\end{split}
\end{equation}
and
\begin{equation}\label{eqn:wns_</>}
	\begin{split}
		\widetilde{W}_S^{ns,</>} \left(\omega,m,n\right) =	\widetilde{\Phi}^{<\>}_S \left(\omega,m,n\right) 
		\\
		+ \left[\widetilde{\Phi}^{R}_S \circ  \widetilde{P}^{R}_S   \circ \widetilde{W}_S^{ns,</>} \right] \left(\omega,m,n\right)
		\\
		+ \left[\widetilde{\Phi}^{R}_S \circ  \widetilde{P}^{</>}_S   \circ \widetilde{W}_S^{ns,A} \right] \left(\omega,m,n\right)
		\\
		+ \left[\widetilde{\Phi}^{</>}_S \circ  \widetilde{P}^{A}_S \circ \widetilde{W}_S^{ns,A} \right] \left(\omega,m,n\right);
	\end{split}
\end{equation}
for the particle-particle T-matrix,
\begin{equation}\label{eq:se_TPP_R/A}
	\begin{split}
		\Sigma^{PP,R/A}_S \left(\omega,n\right) \\
		= i\left[T^{PP,<} \;\square_- \; G^{A/R}_{\bar S} + T^{PP,R/A} \;\square_- \; G^{<}_{\bar S} \right]\left(\omega,n\right),
	\end{split}
\end{equation}
\begin{equation}\label{eq:se_TPP_</>}
	\Sigma^{PP,</>}_S  \left(\omega,n\right) = i\left[T^{PP,</>} \;\square_- \; G^{>/<}_{\bar S} \right]\left(\omega,n\right),
\end{equation}
\begin{equation}\label{eqn:TPP_R/A}
	\begin{split}
		\widetilde{T}^{PP,R/A} \left(\omega,m,n\right) = - U \widetilde{G}^{H,R/A} \left(\omega,m,n\right) U
		\\
		+  U \left[\widetilde{G}^{H,R/A} \circ \widetilde{T}^{PP,R/A} \right] (\omega,m,n),
	\end{split}
\end{equation}
\begin{equation}\label{eqn:TPP_</>}
	\begin{split}
		\widetilde{T}^{PP,</>} \left(\omega,m,n\right) = - U \widetilde{G}^{H,</>} \left(\omega,m,n\right) U
		\\
		+  U \left[\widetilde{G}^{H,R} \circ \widetilde{T}^{PP,</>} \right] (\omega,m,n)
		\\
		+  U \left[\widetilde{G}^{H,</>} \circ \widetilde{T}^{PP,A} \right] (\omega,m,n),
	\end{split}
\end{equation}
\begin{equation}\label{eq:G_H_R/A}
	\begin{split}
		G^{H,R/A} (\omega,n) \\ = i \left[ G^<_A \;\square_+\;  G^{R/A}_B +  G^{R/A}_A \; \square_+ \; G^{>}_B \right], (\omega,n),
	\end{split}
\end{equation}
and
\begin{equation}\label{eq:G_H_</>}
	\begin{split}
		G^{H,</>} (\omega,n) \\ = i \left[ G^{</>}_A \;\square_+\;  G^{</>}_B \right] (\omega,n);
	\end{split}
\end{equation}
and for the particle-hole T-matrix,
\begin{equation}\label{eq:se_TPH_R/A}
	\begin{split}
		\Sigma^{PH,R/A}_S \left(\omega,n\right) \\
		= i\left[T_S^{PH,<} \;\square_+ \; G^{R/A}_{\bar S} + T_S^{PH,R/A} \;\square_+ \; G^{>}_{\bar S} \right]\left(\omega,n\right),
	\end{split}
\end{equation}
\begin{equation}\label{eq:se_TPH_</>}
	\Sigma^{PH,</>}_S  \left(\omega,n\right) = i\left[T_S^{PH,</>} \;\square_+ \; G^{</>}_{\bar S} \right]\left(\omega,n\right),
\end{equation}
\begin{equation} \label{eqn:TPH_R/A}
	\begin{split}
		\widetilde{T}_S^{PH,R/A} \left(\omega,m,n\right) = U \widetilde{G}_{S\bar S}^{F,R/A} \left(\omega,m,n\right) U
		\\
		-  U \left[\widetilde{G}_{S\bar S}^{F,R/A} \circ \widetilde{T}^{PH,R/A}_{S} \right] (\omega,m,n),
	\end{split}
\end{equation}
\begin{equation} \label{eqn:TPH_</>}
	\begin{split}
		\widetilde{T}_S^{PH,</>} \left(\omega,m,n\right) = U \widetilde{G}_{S\bar S}^{F,</>} \left(\omega,m,n\right) U
		\\
		-  U \left[\widetilde{G}_{S\bar S}^{F,</>} \circ \widetilde{T}^{PH,A}_{S} \right] (\omega,m,n)
		\\
		-  U \left[\widetilde{G}_{S\bar S}^{F,R} \circ \widetilde{T}^{PH,</>}_{S} \right] (\omega,m,n),
	\end{split}
\end{equation}
\begin{equation}\label{eq:G_F_R/A}
	\begin{split}
		G_{S\bar S}^{F,R/A} (\omega,n) \\ = -i \left[ G^<_S \;\square_-\;  G^{A/R}_{\bar S} +  G^{R/A}_S \; \square_- \; G^{<}_{\bar S} \right] (\omega,n),
	\end{split}
\end{equation}
and
\begin{equation}\label{eq:G_F_</>}
	\begin{split}
		G_{S\bar S}^{F,</>} (\omega,n) \\ = -i \left[ G^{</>}_S \;\square_-\;  G^{>/<}_{\bar S} \right] (\omega,n).
	\end{split}
\end{equation}

\begin{figure*}[pt!]
	\centering
	\hspace*{-4.5cm} 
	\begin{subfigure}[]{1in}
		\centering
		\includegraphics[width=2.7\textwidth]{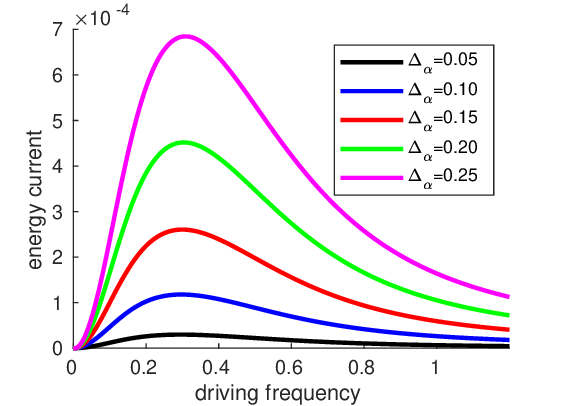} 
		\caption{\label{fig:plotB1}}
	\end{subfigure}
	\qquad\qquad\qquad\qquad\qquad\qquad
	\begin{subfigure}[]{1in}
		\centering
		\includegraphics[width=2.7\textwidth]{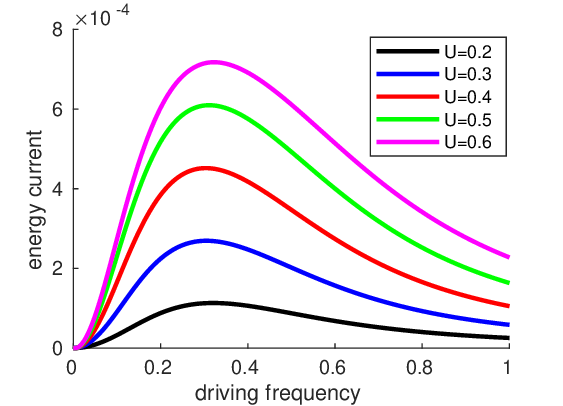} 
		\caption{\label{fig:plotB2}}
	\end{subfigure}
	\qquad\qquad\qquad\qquad\qquad\qquad
	\begin{subfigure}[]{1in}
		\centering
		\includegraphics[width=2.7\textwidth]{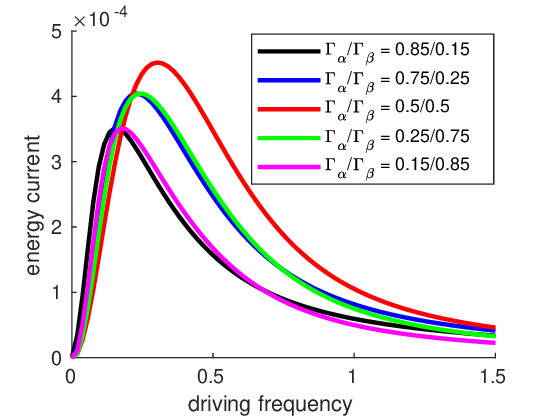} 
		\caption{\label{fig:plotB3}}
	\end{subfigure}
	
	\caption{Time-averaged energy current through the system due to the periodic driving of the left lead. The parameters, unless specified, are $\Gamma_\alpha=\Gamma_\beta=0.5$, $\epsilon_A=\epsilon_B$=-0.2, $U=0.4$, $T=0.001$, $\mu_\alpha=\mu_\beta=0$ and $\Delta_\alpha=0.2$. The discretization was taken at $0.01$, the bounds of integration between $-40$ and $40$, and 49 Fourier coefficients were used. The convergence for both dots was taken as $10^{-4}$}
	\label{fig:plotB}
\end{figure*}

The time-dependent driving of the leads' energies was taken as sinusoidal, 
\begin{equation}\label{eq:phase_definition}
	\psi_{\alpha} (t) = \Delta_\alpha \cos (\Omega t),
\end{equation}
giving $\Psi_\alpha(t)=\left(\Delta_\alpha / \Omega \right)\sin \left(\Omega t\right)$, which can be expanded with the Jacobi-Anger expansion
\begin{equation}\label{eq:jacobi_anger}
	e^{i \frac{\Delta_\alpha}{\Omega}\sin(\Omega t)} = \sum_{n=-\infty}^{n=\infty} J_n \left(\frac{\Delta_\alpha}{\Omega}\right)  e^{in\Omega t},
\end{equation} 
where $J_n(x)$ are Bessel functions of the first kind. We can recast equation  (\ref{eq:lead_self_energies}), as a Floquet matrices,
\begin{equation}
	\bar{\Sigma}_{\sigma} \left(\omega,m,n\right) = \left[ \widetilde{S}_{\sigma} \circ {\widetilde{\Sigma}'}_{\sigma} \circ {\widetilde{S}_{\sigma}}^\dagger \right] \left(\omega,m,n\right),
\end{equation}
where $\widetilde{S}_{\sigma}\left(m,n\right) = J_{m-n} \left(\Delta_\sigma / \Omega \right)$.

In a similar manner to the equations of motion, the observables can be cast in terms of Fourier coefficients: 
\begin{equation}
	\begin{split}
		I^P_\sigma(n-m) = \widetilde{I}^P_{\sigma}\left(m,n\right) \\
		=  2 \int^{\infty}_{-\infty} \frac{d\omega}{2\pi} \; \left[ \widetilde{G}_{S}^R \circ \widetilde{\Sigma}^<_{\sigma}  + \widetilde{G}_{S}^< \circ \widetilde{\Sigma}^A_{\sigma}  \right] (\omega,m,n),
	\end{split}
\end{equation}
\begin{equation}\label{eq:electronic_occupation_floquet}
	\begin{split}
		n_{S} \left(n-m\right) = \widetilde{n}_{S} \left(m,n\right) = -i \int^{\infty}_{-\infty} \frac{d\omega}{2\pi}  \widetilde{G}_{S}^< (\omega,m,n),
	\end{split}
\end{equation}
and
\begin{equation}
	\begin{split}
		I^E_\sigma(n-m) = \widetilde{I}^E_{\sigma}\left(m,n\right) \\
		= - 2 \int^{\infty}_{-\infty} \frac{d\omega}{2\pi} \; \left(\omega + m\Omega \right) \left[\widetilde{\Sigma}_{\sigma}^< \circ \widetilde{G}^A_{S} + \widetilde{\Sigma}_{\sigma}^R \circ \widetilde{G}^<_{S} \right] (\omega,m,n).
	\end{split}
\end{equation}

\subsection{Implementation}

To solve the equations of motion, we invert equations (\ref{eqn:KB-R/A}),(\ref{eqn:wns_R/A}),(\ref{eqn:wns_</>}),(\ref{eqn:TPP_R/A}),(\ref{eqn:TPP_</>}),(\ref{eqn:TPH_R/A}) and (\ref{eqn:TPH_</>}) by first truncating the Floquet matrices, as defined in Eq. (\ref{eqn:floquet_matrix}). Equations (\ref{eq:se_GW_R/A}), (\ref{eq:se_GW_</>}), (\ref{eq:P_s_R/A}), (\ref{eq:P_s_</>}),(\ref{eq:se_TPP_R/A}), (\ref{eq:se_TPP_</>}), (\ref{eq:G_H_R/A}), (\ref{eq:G_H_</>}), (\ref{eq:se_TPH_R/A}), (\ref{eq:se_TPH_</>}), (\ref{eq:G_F_R/A}) and (\ref{eq:G_F_</>}) are calculated by using the Fourier coefficient taken from the appropriate Floquet matrices. These Fourier coefficients were taken from the terms of the Floquet matrices given by  $n+m=0,-1$ of equation (\ref{eqn:floquet_matrix}). The self-energy terms were then transformed back to Floquet matrix form, using equation (\ref{eqn:floquet_matrix}), for use in the equations of motion. The self-consistent process begins with calculating the noninteracting case followed by the interaction self-energies, as specified above. The interaction self-energies are then used to calculate successive iterations of the Green's functions before the following convergence is satisfied:
\begin{equation}
	\begin{split}
		\frac{\sum_{m} \left| n_m^{k} - n_m^{k-1} \right|}{\sum_{m} \left|n^{k}_m\right|} \leq \delta,
	\end{split}
	\end {equation}
where $n^k_m$ is the $k$th iteration of the $m$th Fourier coefficent of the occupation in question, with $\delta$ as the convergence. This convergence was satisfied for each dot's occupation.

\begin{figure*}[pt!]
	\centering
	\hspace*{-4.5cm} 
	\begin{subfigure}[]{1in}
		\centering
		\includegraphics[width=2.7\textwidth]{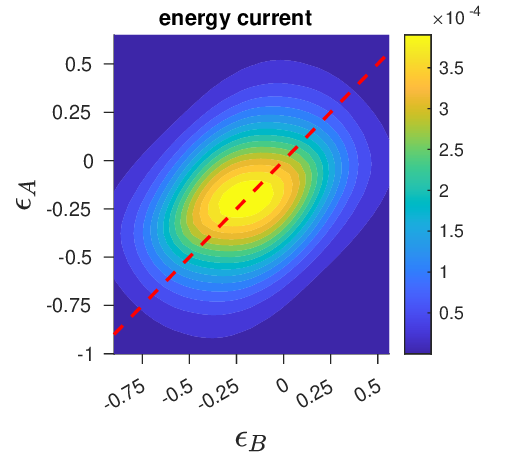} 
		\caption{\label{fig:plotC1}}
	\end{subfigure}
	\qquad\qquad\qquad\qquad\qquad\qquad
	\begin{subfigure}[]{1in}
		\centering
		\includegraphics[width=2.7\textwidth]{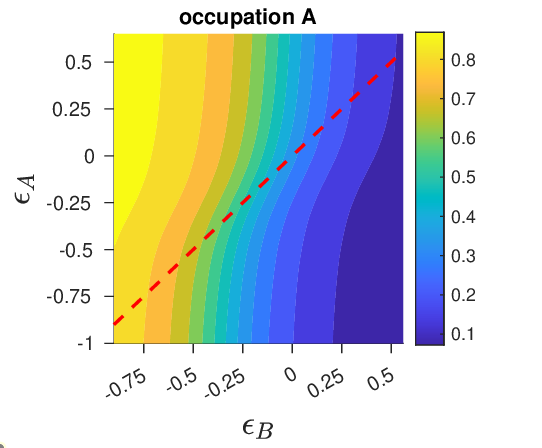} 
		\caption{\label{fig:plotC2}}
	\end{subfigure}
	\qquad\qquad\qquad\qquad\qquad\qquad
	\begin{subfigure}[]{1in}
		\centering
		\includegraphics[width=2.7\textwidth]{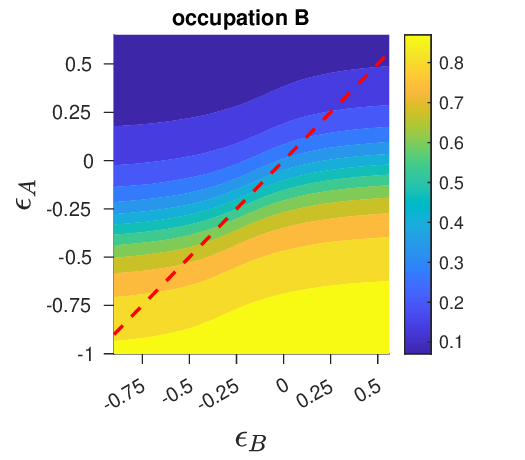} 
		\caption{\label{fig:plotC3}}
	\end{subfigure}
	
	\caption{Time-averaged observables through the system, with driving of the left lead. The parameters are $\Gamma_\alpha=\Gamma_\beta=0.5$, $U=0.4$, $T=0.001$, $\mu_\alpha=\mu_\beta=0$, $\Delta_\alpha=0.2$ and $\Omega = 0.4$. The discretization was taken at $0.01$, the bounds of integration between $-40$ and $40$ and 49 Fourier coefficients were used. The convergence for both dots was taken as $10^{-4}$. The red dashed line follows $\epsilon_A=\epsilon_B$.}
	\label{fig:plotC}
\end{figure*}

\section{Results and Discussion}
\label{sec:discussion}

Within the system, energy moves from the driven to the undriven section via the capacitive coupling of the two dots. In particular, energy transfer occurs when the driven dot is occupied at a higher energy and unoccupied at a lower energy, and the undriven dot is unoccupied at a higher energy and occupied at a lower energy. This process is complicated because the energies at which a dot is occupied are informed by the occupation of the opposing dot. This relationship is transparent in the Hartree approximation, where the average energy current into the central region due to current into the dot $S$ is given by
\begin{equation}\label{eq:avg_eng_current_hartree}
	\bar{I^E_\sigma} = U \int^P_0 \frac{dt}{P} \; n_{\bar S} (t) I^P_\sigma (t).
\end{equation}

These observations, coupled with the sinusoidal nature of the driving, suggest the following approximate cyclic stages in the energy transfer process:
\begin{enumerate} 
	\item Following stage $4$, charge moves onto the driven dot while the undriven dot is largely occupied.
	\item The driven dot is largely occupied as charge moves off the undriven dot.
	\item Charge moves off the driven dot as the undriven dot is largely unoccupied.
	\item The driven dot is largely unoccupied as charge moves onto the undriven dot.
\end{enumerate}
Stages one and two capture the movement of higher energy electrons moving onto the driven dot and off the undriven dot, resulting in the energy transfer from the driven to the undriven region. Stages three and four capture the lower energy electrons moving off the driven dot and onto the undriven dot, resulting in a lower energy transfer than the first two steps in the cycle in the opposite direction. An example of this can be seen in Fig. \ref{fig:plotA}, where the regions in which the stages are most prominent have been highlighted.  

The amount of energy transferred through the system is sensitive to the driving frequency, with the maximum transference a result of the balancing of stages of energy transfer [see in Fig. \ref{fig:plotB}]. As the driving frequency decreases, electrons move between the dots and their respective leads quicker than the energy transfer stages can complete. In particular, the dots that remain largely occupied in stages one and two and largely unoccupied in processes three and four begin to change in occupation, resulting in less pronounced changes in the opposing dot's occupation energy and outgoing energy current. Conversely, as the driving frequency increases, the charge has less time to move between the dots and their respective leads, resulting in smaller maxima and minima for the occupations over the period, which reduces the opposing dot's occupation energy and its outgoing energy current.

The driving profile and the system parameters beyond the driving frequency also inform the effectiveness of the energy transfer process. As expected, given Eq. (\ref{eq:avg_eng_current_hartree}), increases in the interaction strength $U$ were found to increase the average energy current through the system [see Fig. \ref{fig:plotB2}]. It was also found that significant asymmetry of the coupling strengths diminished energy flow through the system [see Fig. \ref{fig:plotB3}]. This is due to uneven transfer rates, $\tau_\sigma \sim 1 / \Gamma_\sigma$, for the movement of electrons between the dots and their respective leads, resulting in the inefficient completion of the stages of the energy transfer process. For the energies of the dots, the largest transference of energy through the system was achieved with dots with equal energies situated below the chemical potential of the two leads, such that, on average, the dots are both around half filled [see Fig. \ref{fig:plotC}].

\section{Conclusion} 
\label{ref:conclusion}

We have investigated two capacitively coupled quantum dots coupled to respective leads, where one lead's energies are driven sinusoidally. While particles cannot move between the dots, Coulomb repulsion between the dots allows for the transfer of energy. The stages of the energy transfer were identified, and the effects of system parameters' were investigated. In particular, it was found that energy transfer was maximized for a given driving, corresponding to the efficient completion of the identified energy transfer process.

This work has focused on a regime of relatively weak coupling with $\Gamma > U$. Further work in a regime of $\Gamma < U$ may result in different energy transfer stages, as outlined in Sec. \ref{sec:discussion}, for various drivings, suggesting interesting possible avenues for further research. Moreover, more complicated driving profiles and statistics relating to energy transfer may prove valuable in understanding and manipulating energy transfer in systems like that investigated.

This result furthers the understanding of particle and energy transfer in capacitively coupled quantum dots, particularly within the context of nonadiabatic driving. This is particularly important as the miniaturization of nanoelectronics brings active elements closer together, resulting in the potential for unwarranted capacitive coupling. 

\newpage

\bibliographystyle{unsrt}
\bibliography{references}

\end{document}